\documentclass[aps,prd,twocolumn,showpacs,superscriptaddress,amsmath,amssymb]{revtex4-1}

\usepackage{graphicx}
\usepackage{dcolumn}
\usepackage{bm}
\usepackage{hyperref}
\usepackage{verbatim}

\newcommand{\ee}[1]{\!\times\!10^{#1}}
\newcommand{\gws}{gravitational-waves }
\newcommand{\gw}{gravitational-wave }

\newcommand{\beq}{\begin{equation}}
\newcommand{\eeq}{\end{equation}}

\def\be{\begin{equation}}
\def\ee{\end{equation}}

\begin{document}

\title{Narrow-band search of continuous gravitational-wave signals from Crab and Vela pulsars in Virgo VSR4 data}




\author{%
J.~Aasi,$^{1}$
B.~P.~Abbott,$^{1}$
R.~Abbott,$^{1}$
T.~Abbott,$^{2}$
M.~R.~Abernathy,$^{1}$
F.~Acernese,$^{3,4}$
K.~Ackley,$^{5}$
C.~Adams,$^{6}$
T.~Adams,$^{7,8}$
T.~Adams,$^{8}$
% ...
}\noaffiliation

\affiliation {LIGO, California Institute of Technology, Pasadena, CA 91125, USA }
\affiliation {Louisiana State University, Baton Rouge, LA 70803, USA }
\affiliation {Universit\`a di Salerno, Fisciano, I-84084 Salerno, Italy }
\affiliation {INFN, Sezione di Napoli, Complesso Universitario di Monte S.Angelo, I-80126 Napoli, Italy }
\affiliation {University of Florida, Gainesville, FL 32611, USA }
\affiliation {LIGO Livingston Observatory, Livingston, LA 70754, USA }
\affiliation {Cardiff University, Cardiff, CF24 3AA, United Kingdom }
\affiliation {Laboratoire d'Annecy-le-Vieux de Physique des Particules (LAPP), Universit\'e de Savoie, CNRS/IN2P3, F-74941 Annecy-le-Vieux, France }
\affiliation {University of Sannio at Benevento, I-82100 Benevento, Italy and INFN, Sezione di Napoli, I-80100 Napoli, Italy }
\affiliation {Experimental Group, Albert-Einstein-Institut, Max-Planck-Institut f\"ur Gravi\-ta\-tions\-physik, D-30167 Hannover, Germany }
\affiliation {Nikhef, Science Park, 1098 XG Amsterdam, The Netherlands }
\affiliation {LIGO, Massachusetts Institute of Technology, Cambridge, MA 02139, USA }
\affiliation {Instituto Nacional de Pesquisas Espaciais, 12227-010 - S\~{a}o Jos\'{e} dos Campos, SP, Brazil }
\affiliation {Inter-University Centre for Astronomy and Astrophysics, Pune - 411007, India }
\affiliation {International Centre for Theoretical Sciences, Tata Institute of Fundamental Research, Bangalore 560012, India }
\affiliation {Syracuse University, Syracuse, NY 13244, USA }
\affiliation {Data Analysis Group, Albert-Einstein-Institut, Max-Planck-Institut f\"ur Gravitations\-physik, D-30167 Hannover, Germany }
\affiliation {University of Wisconsin--Milwaukee, Milwaukee, WI 53201, USA }
\affiliation {Universit\`a di Siena, I-53100 Siena, Italy }
\affiliation {INFN, Sezione di Pisa, I-56127 Pisa, Italy }
\affiliation {The University of Mississippi, University, MS 38677, USA }
\affiliation {California State University Fullerton, Fullerton, CA 92831, USA }
\affiliation {University of Southampton, Southampton, SO17 1BJ, United Kingdom }
\affiliation {Leibniz Universit\"at Hannover, D-30167 Hannover, Germany }
\affiliation {INFN, Sezione di Roma, I-00185 Roma, Italy }
\affiliation {University of Birmingham, Birmingham, B15 2TT, United Kingdom }
\affiliation {Albert-Einstein-Institut, Max-Planck-Institut f\"ur Gravitations\-physik, D-14476 Golm, Germany }
\affiliation {Montana State University, Bozeman, MT 59717, USA }
\affiliation {Universit\`a di Perugia, I-06123 Perugia, Italy }
\affiliation {INFN, Sezione di Perugia, I-06123 Perugia, Italy }
\affiliation {European Gravitational Observatory (EGO), I-56021 Cascina, Pisa, Italy }
\affiliation {SUPA, University of Glasgow, Glasgow, G12 8QQ, United Kingdom }
\affiliation {LIGO Hanford Observatory, Richland, WA 99352, USA }
\affiliation {APC, AstroParticule et Cosmologie, Universit\'e Paris Diderot, CNRS/IN2P3, CEA/Irfu, Observatoire de Paris, Sorbonne Paris Cit\'e, 10, rue Alice Domon et L\'eonie Duquet, F-75205 Paris Cedex 13, France }
\affiliation {Columbia University, New York, NY 10027, USA }
\affiliation {Stanford University, Stanford, CA 94305, USA }
\affiliation {Universit\`a di Pisa, I-56127 Pisa, Italy }
\affiliation {CAMK-PAN, 00-716 Warsaw, Poland }
\affiliation {Astronomical Observatory Warsaw University, 00-478 Warsaw, Poland }
\affiliation {The University of Texas at Brownsville, Brownsville, TX 78520, USA }
\affiliation {Universit\`a degli Studi di Genova, I-16146 Genova, Italy }
\affiliation {INFN, Sezione di Genova, I-16146 Genova, Italy }
\affiliation {RRCAT, Indore MP 452013, India }
\affiliation {Faculty of Physics, Lomonosov Moscow State University, Moscow 119991, Russia }
\affiliation {LAL, Universit\'e Paris-Sud, IN2P3/CNRS, F-91898 Orsay, France }
\affiliation {NASA/Goddard Space Flight Center, Greenbelt, MD 20771, USA }
\affiliation {University of Western Australia, Crawley, WA 6009, Australia }
\affiliation {Department of Astrophysics/IMAPP, Radboud University Nijmegen, P.O. Box 9010, 6500 GL Nijmegen, The Netherlands }
\affiliation {ARTEMIS, Universit\'e Nice-Sophia-Antipolis, CNRS and Observatoire de la C\^ote d'Azur, F-06304 Nice, France }
\affiliation {MTA E\"otv\"os University, `Lendulet' A. R. G., Budapest 1117, Hungary }
\affiliation {Institut de Physique de Rennes, CNRS, Universit\'e de Rennes 1, F-35042 Rennes, France }
\affiliation {Washington State University, Pullman, WA 99164, USA }
\affiliation {Universit\`a degli Studi di Urbino 'Carlo Bo', I-61029 Urbino, Italy }
\affiliation {INFN, Sezione di Firenze, I-50019 Sesto Fiorentino, Firenze, Italy }
\affiliation {University of Oregon, Eugene, OR 97403, USA }
\affiliation {Laboratoire Kastler Brossel, ENS, CNRS, UPMC, Universit\'e Pierre et Marie Curie, F-75005 Paris, France }
\affiliation {VU University Amsterdam, 1081 HV Amsterdam, The Netherlands }
\affiliation {University of Maryland, College Park, MD 20742, USA }
\affiliation {University of Massachusetts Amherst, Amherst, MA 01003, USA }
\affiliation {Laboratoire des Mat\'eriaux Avanc\'es (LMA), IN2P3/CNRS, Universit\'e de Lyon, F-69622 Villeurbanne, Lyon, France }
\affiliation {Universitat de les Illes Balears---IEEC, E-07122 Palma de Mallorca, Spain }
\affiliation {Universit\`a di Napoli 'Federico II', Complesso Universitario di Monte S.Angelo, I-80126 Napoli, Italy }
\affiliation {Canadian Institute for Theoretical Astrophysics, University of Toronto, Toronto, Ontario, M5S 3H8, Canada }
\affiliation {Tsinghua University, Beijing 100084, China }
\affiliation {University of Michigan, Ann Arbor, MI 48109, USA }
\affiliation {INFN, Sezione di Roma Tor Vergata, I-00133 Roma, Italy }
\affiliation {National Tsing Hua University, Hsinchu Taiwan 300 }
\affiliation {Charles Sturt University, Wagga Wagga, NSW 2678, Australia }
\affiliation {Caltech-CaRT, Pasadena, CA 91125, USA }
\affiliation {Pusan National University, Busan 609-735, Korea }
\affiliation {Australian National University, Canberra, ACT 0200, Australia }
\affiliation {Carleton College, Northfield, MN 55057, USA }
\affiliation {Universit\`a di Roma Tor Vergata, I-00133 Roma, Italy }
\affiliation {INFN, Gran Sasso Science Institute, I-67100 L'Aquila, Italy }
\affiliation {Universit\`a di Roma 'La Sapienza', I-00185 Roma, Italy }
\affiliation {University of Brussels, Brussels 1050 Belgium }
\affiliation {Sonoma State University, Rohnert Park, CA 94928, USA }
\affiliation {Texas Tech University, Lubbock, TX 79409, USA }
\affiliation {University of Minnesota, Minneapolis, MN 55455, USA }
\affiliation {The University of Sheffield, Sheffield S10 2TN, United Kingdom }
\affiliation {Wigner RCP, RMKI, H-1121 Budapest, Konkoly Thege Mikl\'os \'ut 29-33, Hungary }
\affiliation {Montclair State University, Montclair, NJ 07043, USA }
\affiliation {Argentinian Gravitational Wave Group, Cordoba Cordoba 5000, Argentina }
\affiliation {Universit\`a di Trento, I-38123 Povo, Trento, Italy }
\affiliation {INFN, Trento Institute for Fundamental Physics and Applications, I-38123 Povo, Trento, Italy }
\affiliation {Northwestern University, Evanston, IL 60208, USA }
\affiliation {University of Cambridge, Cambridge, CB2 1TN, United Kingdom }
\affiliation {University of Szeged, D\'om t\'er 9, Szeged 6720, Hungary }
\affiliation {Rutherford Appleton Laboratory, HSIC, Chilton, Didcot, Oxon, OX11 0QX, United Kingdom }
\affiliation {Embry-Riddle Aeronautical University, Prescott, AZ 86301, USA }
\affiliation {The Pennsylvania State University, University Park, PA 16802, USA }
\affiliation {American University, Washington, DC 20016, USA }
\affiliation {University of Adelaide, Adelaide, SA 5005, Australia }
\affiliation {West Virginia University, Morgantown, WV 26506, USA }
\affiliation {Raman Research Institute, Bangalore, Karnataka 560080, India }
\affiliation {Korea Institute of Science and Technology Information, Daejeon 305-806, Korea }
\affiliation {University of Bia{\l }ystok, 15-424 Bia{\l }ystok, Poland }
\affiliation {SUPA, University of Strathclyde, Glasgow, G1 1XQ, United Kingdom }
\affiliation {IISER-TVM, CET Campus, Trivandrum Kerala 695016, India }
\affiliation {Institute of Applied Physics, Nizhny Novgorod, 603950, Russia }
\affiliation {Seoul National University, Seoul 151-742, Korea }
\affiliation {Hanyang University, Seoul 133-791, Korea }
\affiliation {NCBJ, 05-400 \'Swierk-Otwock, Poland }
\affiliation {IM-PAN, 00-956 Warsaw, Poland }
\affiliation {Institute for Plasma Research, Bhat, Gandhinagar 382428, India }
\affiliation {The University of Melbourne, Parkville, VIC 3010, Australia }
\affiliation {INFN, Sezione di Padova, I-35131 Padova, Italy }
\affiliation {Monash University, Victoria 3800, Australia }
\affiliation {ESPCI, CNRS, F-75005 Paris, France }
\affiliation {Universit\`a di Camerino, Dipartimento di Fisica, I-62032 Camerino, Italy }
\affiliation {Southern University and A\&M College, Baton Rouge, LA 70813, USA }
\affiliation {College of William and Mary, Williamsburg, VA 23187, USA }
\affiliation {Abilene Christian University, Abilene, TX 79699, USA }
\affiliation {Instituto de F\'\i sica Te\'orica, Univ. Estadual Paulista/ICTP South American Institute for Fundamental Research, S\~ao Paulo SP 01140-070, Brazil }
\affiliation {IISER-Kolkata, Mohanpur, West Bengal 741252, India }
\affiliation {Whitman College, 280 Boyer Ave, Walla Walla, WA 9936, USA }
\affiliation {National Institute for Mathematical Sciences, Daejeon 305-390, Korea }
\affiliation {Rochester Institute of Technology, Rochester, NY 14623, USA }
\affiliation {Hobart and William Smith Colleges, Geneva, NY 14456, USA }
\affiliation {Tata Institute for Fundamental Research, Mumbai 400005, India }
\affiliation {SUPA, University of the West of Scotland, Paisley, PA1 2BE, United Kingdom }
\affiliation {Institute of Astronomy, 65-265 Zielona G\'ora, Poland }
\affiliation {Universit\"at Hamburg, D-22761 Hamburg, Germany }
\affiliation {Indian Institute of Technology, Gandhinagar Ahmedabad Gujarat 382424, India }
\affiliation {Andrews University, Berrien Springs, MI 49104, USA }
\affiliation {Trinity University, San Antonio, TX 78212, USA }
\affiliation {University of Washington, Seattle, WA 98195, USA }


\begin{abstract}
In this paper we present the results of a coherent narrow-band search for continuous gravitational-wave signals from the Crab and Vela pulsars conducted on Virgo VSR4 data. In order to take into account a possible small mismatch between the gravitational wave frequency and two times the star rotation frequency, inferred from measurement of the electromagnetic pulse rate, a range of 0.02 Hz around two times the star rotational frequency has been searched for both the pulsars. No evidence for a signal has been found and 95$\%$ confidence level upper limits have been computed both assuming polarization parameters are completely unknown and that they are known with some uncertainty, as derived from X-ray observations of the pulsar wind torii. For Vela the upper limits are comparable to the spin-down limit, computed assuming that all the observed spin-down is due to the emission of gravitational waves. For Crab the upper limits are about a factor of two below the spin-down limit, and represent a significant improvement with respect to past analysis. This is the first time the spin-down limit is significantly overcome in a narrow-band search.  
\end{abstract}


\maketitle

\section{Introduction}
Continuous gravitational-wave signals (CW) are emitted by spinning neutron stars if asymmetric with respect to the rotation axis. The asymmetry can be due to various mechanisms, like a non-axismmetric residual strain from the star's birth or a strong internal magnetic field not aligned to the rotation axis, see e.g. \cite{ref:review}. When the source parameters, position, frequency, spin-down, are known with high accuracy {\it targeted} searches can be done using optimal analysis methods, based on matched filtering. This happens, for instance, with known pulsars for which accurate ephemerides are often obtained from electromagnetic (EM) observations, especially in the radio, gamma-ray and X-ray band.  
This means that a strict correlation between the gravitational wave frequency $f_0$ and the star's measured rotational frequency $f_{rot}$ is assumed. In the classical case of a non-axisymmetric neutron star rotating around one of its principal axes of inertia the gravitational frequency would be exactly twice the rotation frequency of the star. Several targeted searches for CW have been conducted on data from first generation interferometric detectors. No evidence for a signal has been found, but interesting upper limits have been placed in a few cases \cite{ref:s5_kp,ref:vela_vsr2,ref:vsr4_kp}. Given the uncertainties on gravitational wave emission mechanisms and also the lack of a full detailed picture of the electro-magnetic emission geometry, it is not obvious at all that the gravitational-wave emission takes place at {\it exactly} twice the star measured pulse rate, or that such relation holds for observation times of months to years. For instance, if a neutron star is made of a crust and a core rotating at slightly different rates, and if the gravitational-wave emission is dominated by an asymmetry in the core then a search targeted at 2$f_{rot}$ would assume a wrong signal frequency.   
We then consider here that the signal frequency can be slightly different with respect to $f_0=2f_{rot}$ and can vary in the interval 
\begin{equation}
f(t) \in \left[f_0(t)(1-\delta),~ f_0(t)(1+\delta)\right ] 
\label{eq:widef}
\end{equation}  
where $\delta$ is a small positive real number. Following the discussion in \cite{ref:s5_nb}, if the star crust and core are linked by some torque that tends to enforce corotation on a timescale $\tau_c$, then $\delta \sim \tau_c /\tau_{sd}$, where $\tau_{sd} \sim f_0/\dot{f}_0$ is the characteristic spin-down time. A relation of the form of Eq.~(\ref{eq:widef}) also holds in the case the gravitational radiation is produced by free precession of a nearly bi-axial star \cite{ref:jones1}, in which case $\delta$ is of the order of $(I_{zz}-I_{xx})/I_{xx}$, where $I_{xx}$ and $I_{zz}$ are the star moments of inertia with respect to a principal axis on the equatorial plane and aligned with the rotation axis respectively. In general, a value of $\delta$ of the order of, say, $10^{-4}$, corresponds to $\tau_c \sim 10^{-4}\tau_{sd}$ which, depending on the specific targeted pulsar can be several months or years. This would be comparable or larger than the longest timescale observed in pulsar glitch recovery where a recoupling between the two components might occur. In terms of free precession, $\delta\sim 10^{-4}$ is on the high end of the range of deformations that neutron stars could sustain \cite{ref:owen,ref:hask,ref:macda}. 

Narrow-band searches have not received much attention until now, one notable exception being the Crab narrow-band search over the first 9 months of LIGO S5 data \cite{ref:s5_nb}, based on the $\mathcal{F}$-statistic \cite{ref:jks}. In previous work \cite{ref:nb_method} an optimal method, based on matched filtering in the space of signal Fourier components,  has been proposed to carry out a search for CW over a small frequency and spin-down range around the values inferred from EM observations. In this paper we describe the application of such a method to a narrow-band search of CW from Crab and Vela pulsars in the data of Virgo VSR4 run. As no evidence for a signal is found, we place upper limits on signal amplitude. 

The outline of the paper is the following. In Sec.~\ref{sec:signal} the expected gravitational wave signal from spinning neutron stars is introduced. In Sec.~\ref{sec:data} the main characteristics of Virgo VSR4 data are presented. In Sec.~\ref{sec:method} the analysis method is described, while in Sec.~\ref{sec:result} the analysis results are discussed. Sec.~\ref{sec:valid} is dedicated to the validation tests of the analysis procedure. Finally, Sec.~\ref{sec:disc} contains the conclusions. 

\section{\label{sec:signal}The gravitational wave signal}
A non-axisymmetric neutron star steadily spinning about one of its principal axis emits a quadrupolar \gw signal at twice the star rotational frequency $f_{rot}$, which at the detector can be described \cite{ref:fivevect} as
\be
h(t)=H_0(H_+A^+(t)+H_{\times}A^{\times}(t))e^{\imath \left(\omega_0(t)t+\Phi_0\right)}
\label{eq:hoft}
\ee
where taking the real part is understood. The constant $\Phi_0$ is the initial signal phase. The angular signal frequency $\omega_0(t)=4\pi f_{rot}(t)$,  is a function of time. Consequently the signal phase  
is not that of a simple monochromatic signal and depends on both the intrinsic rotational frequency and frequency derivatives of 
the pulsar and on Doppler and propagation effects, which include the Einstein delay and, possibly, the Shapiro delay. These variations are corrected in the time-domain in a way described in Sec.~\ref{sec:method}.
The two complex amplitudes $H_+$ and $H_{\times}$ are given respectively by
\begin{equation}
H_+=\frac{\cos{2\psi}-\imath \eta \sin{2\psi}}{\sqrt{1+\eta^2}}
\label{eq:Hp}
\end{equation}
\begin{equation}
H_{\times}=\frac{\sin{2\psi}+\imath \eta \cos{2\psi}}{\sqrt{1+\eta^2}}
\label{eq:Hc}
\end{equation}
in which $\eta$ is the ratio of the polarization ellipse semi-minor to semi-major axis and the polarization angle $\psi$ defines the direction of the major axis with respect to the celestial parallel of the source (measured counterclockwise). 
The functions $A^+$ and $A^{\times}$ describe  the detector response as a function of time and are a linear combination of terms depending on the Earth sidereal angular frequency, $\Omega_{\oplus}$:
$$
A^+ = a_0 + a_{1c} \cos{\Omega_{\oplus} t} + a_{1s} \sin{\Omega_{\oplus} t} +
$$
\begin{equation}
a_{2c} \cos{2\Omega_{\oplus} t} + a_{2s} \sin{2\Omega_{\oplus} t}  
\label{eq:Ap}
\end{equation}

$$
A^{\times} =  b_{1c} \cos{\Omega_{\oplus} t} + b_{1s} \sin{\Omega_{\oplus} t} +
$$
\begin{equation}
b_{2c} \cos{2\Omega_{\oplus} t} + b_{2s} \sin{2\Omega_{\oplus} t}
\label{eq:Ac}
\end{equation}
where the coefficients depend on the source position and detector position and orientation on the Earth \cite{ref:fivevect}. 

As discussed e.g. in \cite{ref:vela_vsr2}, the overall wave amplitude $H_0$ and $\eta$ are related to the ``standard'' signal amplitude 
\begin{equation}
h_0=\frac{4\pi^2G}{c^4}\frac{I_{zz}\varepsilon f^2_0}{d}
\label{eq:h0stand}
\end{equation}
and to the angle $\iota$ between the star rotation axis and the line of sight to the source by 
\begin{equation}
H_0=h_0 \sqrt{\frac{1+6 \cos^2 \iota+ \cos^4 \iota}{4}}
\end{equation} 
\begin{equation}
\eta=-\frac{2 \cos \iota}{1+\cos^2 \iota}. 
\end{equation}
In Eq.~(\ref{eq:h0stand}) $G$ is the gravitational constant, $c$ is the light velocity, $I_{zz}$ is the star moment of inertia with respect to the  
principal axis aligned with the rotation axis, $\varepsilon=\frac{I_{xx}-I_{yy}}{I_{zz}}$ is the equatorial ellipticity expressed in terms of principal moments of inertia, and $d$ is the source distance.

Equating the \gw  luminosity 
to the kinetic energy lost as the pulsar spins down gives us the so-called {\it spin-down limit} on \gw strain
$$
h_0^{\rm sd} = \left(\frac{5}{2} \frac{G I_{zz} \dot{f}_{\rm rot}}{c^3 d^2 f_{\rm rot}}\right)^{1/2}=
$$
\begin{equation}
8.06\times 10^{-19}\frac{I^{1/2}_{38}}{d_{\rm kpc}}\left(\frac{|\dot{f}_{\rm rot}|}{\rm Hz/s}\right)^{1/2}\left(\frac{f_{\rm rot}}{\rm Hz}\right)^{-1/2},
\label{eq:h0sd}
\end{equation}
where $I_{38}$ is the star's moment of inertia in units of $10^{38}$\,kg\,m$^2$, $\dot{f}_{\rm rot}$ is the time derivative of the rotational frequency and $d_{\rm kpc}$ is the 
distance to the pulsar in kiloparsecs. The spin-down limit on the signal amplitude corresponds to an upper limit on the star's fiducial ellipticity
\begin{equation}\label{eq:eps_sd}
\varepsilon^{\rm sd} = 0.237\left(\frac{h_0^{\rm sd}}{10^{-24}}\right)f_{\rm rot}^{-2} I_{38}^{-1} d_{\rm 
kpc}.
\end{equation}
This quantity, for a given neutron star equation of state, can be related to the physical ellipticity of the star surface \cite{ref:macda}. Setting a \gw upper limit below the spin-down limit is an important milestone in CW searches as it allows us to constrain the fraction of spin-down energy due to the emission of \gws, which gives insight into the spin-down energy budget of the pulsar. On the other hand the $l=m=2$ mass quadrupole moment $Q_{22}$ is related to the \gw amplitude by
\begin{equation}\label{eq:q22}
Q_{22} = \sqrt{\frac{15}{8\pi}} I_{zz}\varepsilon = h_0\left( \frac{c^4 d}{16\pi^2G f_{\rm rot}^2} \right) \sqrt{\frac{15}{8\pi}},
\end{equation}
see e.g. \cite{ref:owen}, and is independent of any assumptions about the star's equation of state and its moment of inertia.

\section{\label{sec:data}Instrumental performance in Virgo VSR4 run}
\begin{figure*}[!htbp]
\includegraphics[width=1.0\textwidth]{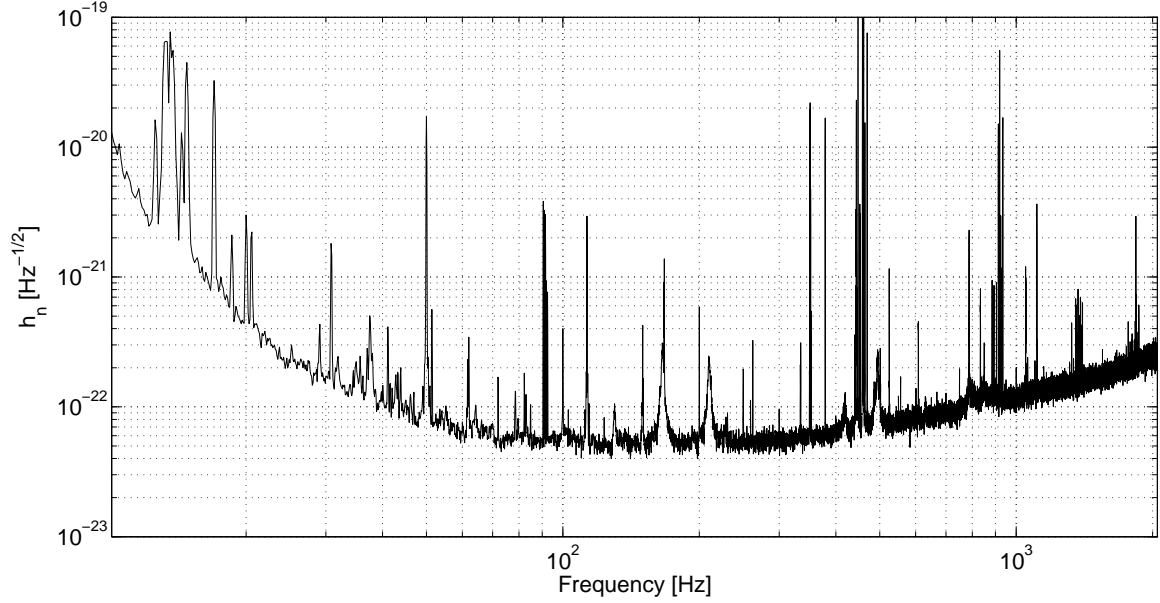}
\caption{A typical sensitivity curve of Virgo VSR4 run, expressed in terms of noise spectral density.
\label{fig:vsr4_sens}}
\end{figure*}

\begin{figure*}[!htbp]
\includegraphics[width=10cm]{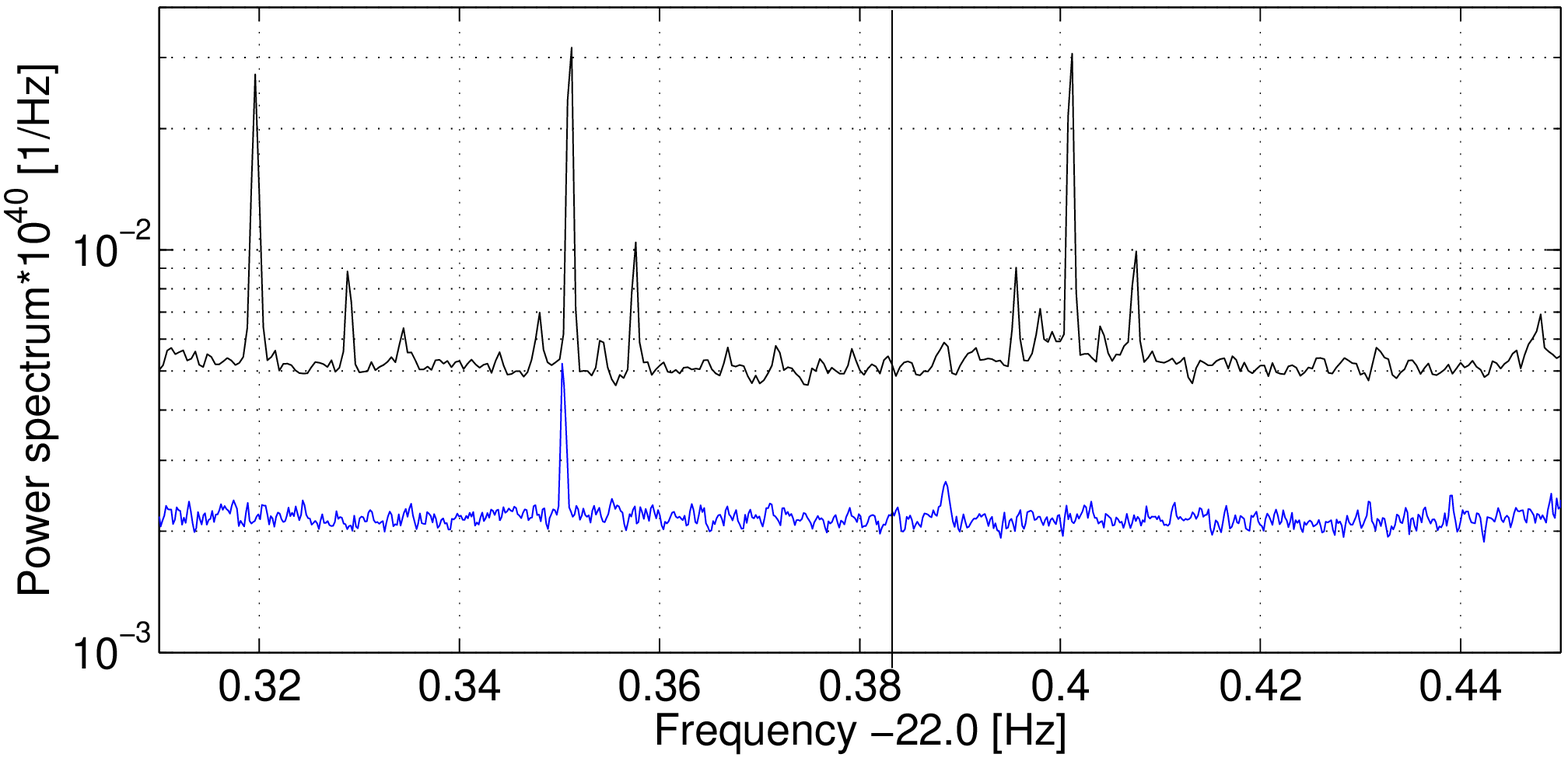}
\includegraphics[width=10cm]{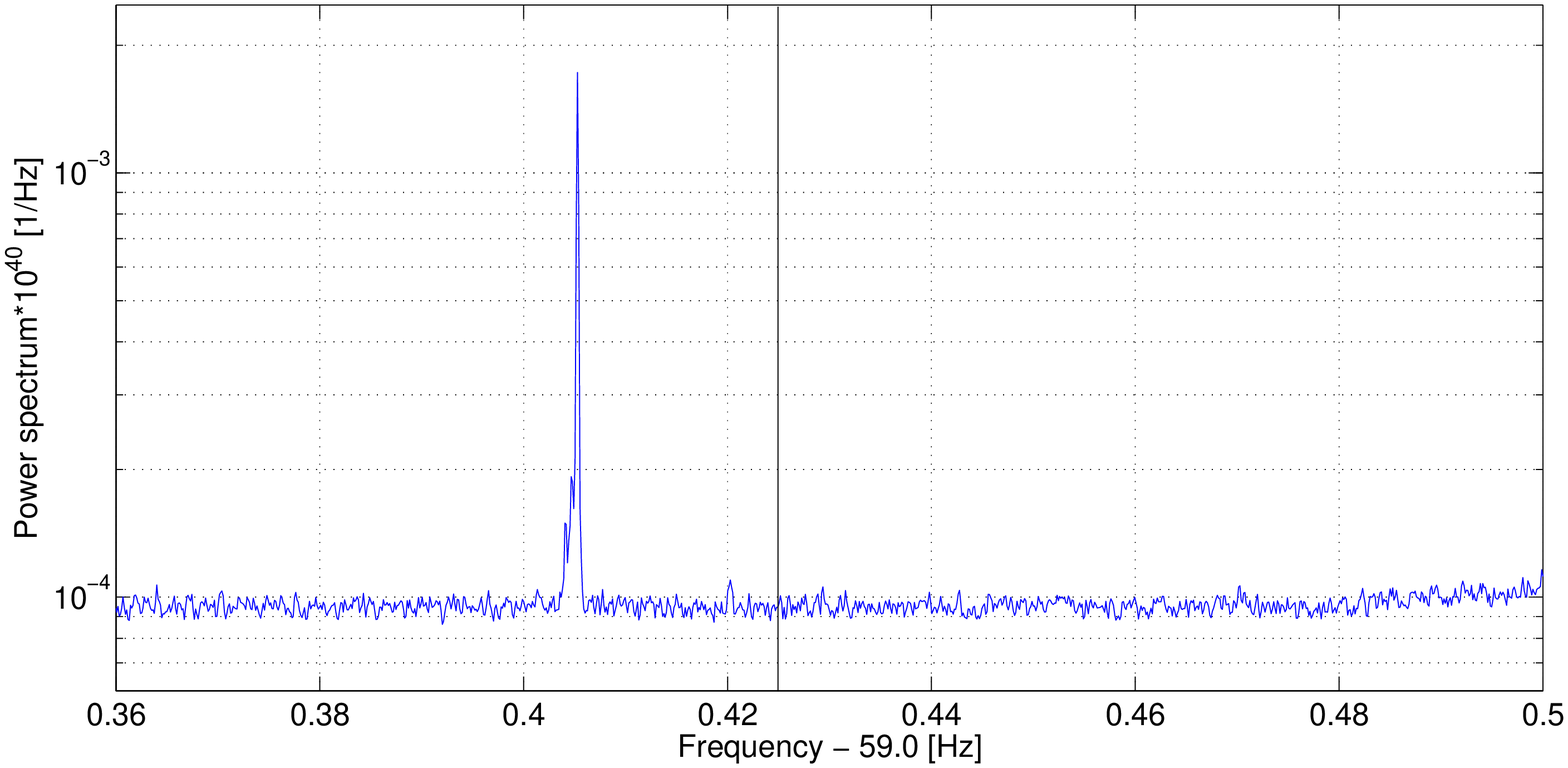}
\caption{Average spectrum of VSR4 data around Vela (top plot) and Crab (bottom plot) reference frequency (identified by the vertical black lines). For Vela two spectra are in fact shown in the same plot. The upper one (black curve) has been computed over the first 29 days of the run when the frequency region around Vela was affected by an instrumental disturbance. This was due to a non-linear coupling between the DARM control line at 379 Hz and calibration lines at 356 Hz and 356.5 Hz and was moved away from the Vela region by shifting of 5 Hertz the frequency of calibration lines. The lower plot (blue curve) is the average spectrum computed after the removal of the disturbance. The large spectral disturbance appearing near the Crab reference frequency is due to a poorly understood coupling between a line at 60 Hz, part of a 10-Hertz comb of lines of likely magnetic origin, and the fundamental pendulum frequency of the Virgo mirror system, at 0.594 Hz.   
\label{fig:crab_vela_ps}}
\end{figure*}

Interferometric gravitational wave detectors, LIGO \cite{ref:ligo}, Virgo \cite{ref:virgo}, GEO \cite{ref:geo}, have collected a large amount of data in recent years.
For the analysis described in this paper we have used calibrated data, sampled at 4096 Hz, from Virgo VSR4 science run. The run extended from June 3rd, 2011 (10:27 UTC) to September, 5th 2011 (13:26 UTC), with a duty factor of about 81$\%$, corresponding to an effective duration of 76 days. Calibration uncertainties amounted to 7.5$\%$ in amplitude and $(40+50f_{kHz})$ mrad in phase up to 500 Hz, where $f_{kHz}$ is the frequency in kilohertz. The uncertainty on the amplitude will contribute to the uncertainty on the upper limit on signal amplitude, together with that coming from the finite size of the Monte Carlo simulation used to compute it, see Sec.~\ref{sec:result}. The calibration error on the phase can be shown to have a negligible impact on the analysis \cite{ref:vela_vsr2}. The low-frequency sensitivity of VSR4 was significantly better than that of previous Virgo runs, especially due to the use of monolithic mirror suspensions, and basically in agreement with the planned sensitivity of the initial Virgo interferometer. In Fig.~\ref{fig:vsr4_sens} a typical VSR4 strain sensitivity curve is shown. 
In Fig.~\ref{fig:crab_vela_ps} the average power spectrum around Crab and Vela reference frequency is plotted. Note the large sensitivity improvement around the Vela frequency after removal of an instrumental disturbance, about one month after the beginning of the run, see the figure caption for more details. 

\section{\label{sec:method}Search description}
\begin{figure*}[!htbp]
\includegraphics[width=10cm]{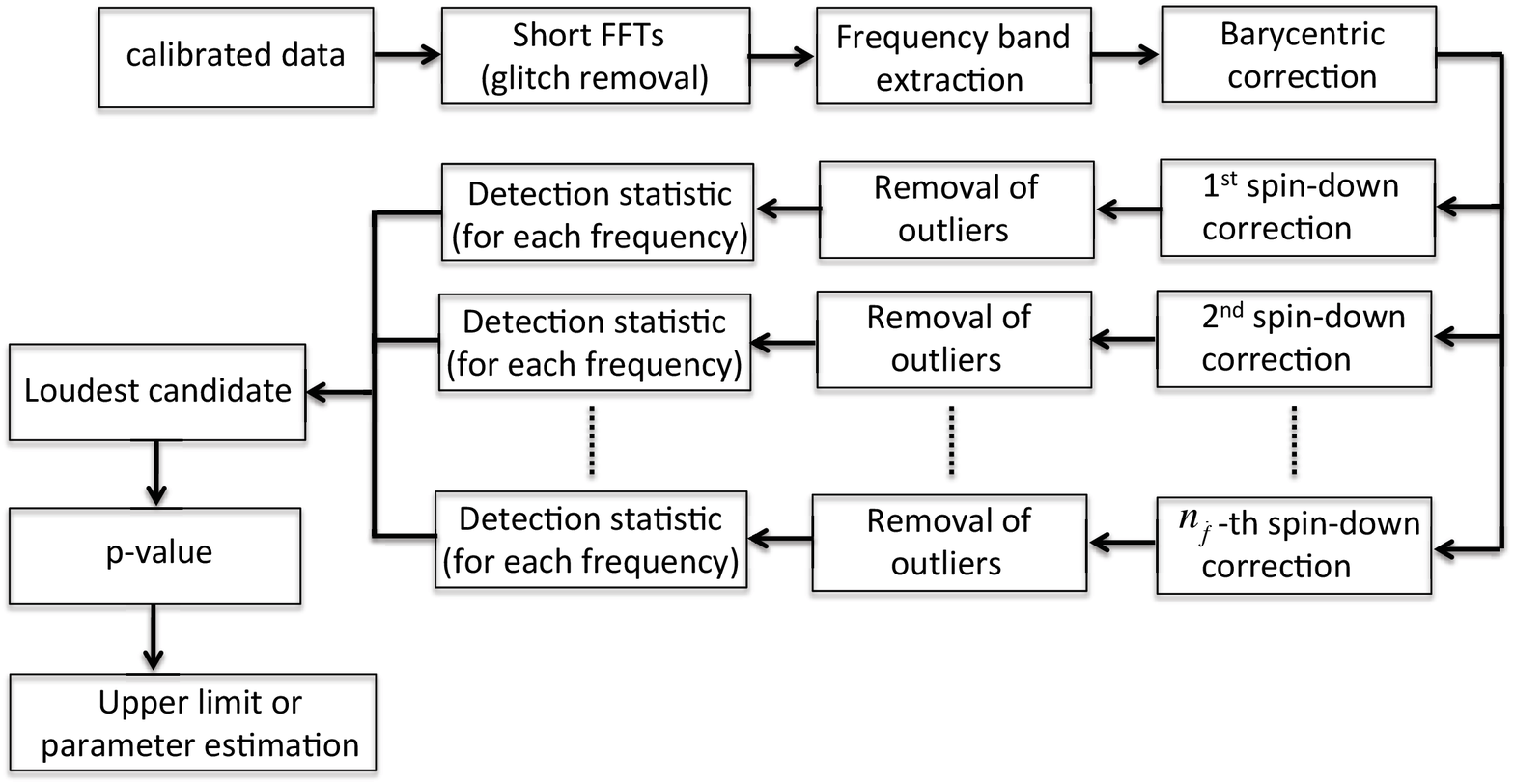}
\caption{Scheme of the narrow-band analysis pipeline. The starting point is constituted by detector calibrated data, sampled at 4096 Hz. After barycentric and spin-down corrections the data are down-sampled at 1 Hz. The number of different spin-down corrections applied to the data, $n_{\dot{f}}$, and then the number of produced corrected time series, is given by Eq.~(\ref{nsd}). See text and \cite{ref:nb_method} for more details.
\label{fig:analysis_scheme}}
\end{figure*}

The analysis pipeline consists of several steps, schematically depicted in Fig.~\ref{fig:analysis_scheme}, which we summarize here. More details are given in \cite{ref:nb_method}. The starting point is a collection of windowed and interlaced (by half) ``short'' (1024 seconds) FFTs (the Short FFT Database - SFDB) built from calibrated detector data \cite{ref:sfdb}. At this stage a first cleaning procedure is applied to the data in order to remove big and short time duration disturbances, that could reduce the search sensitivity. A small frequency band is then extracted from the SFDB, but large enough to contain the Doppler and spin-down variations of the signal possibly emitted by the target pulsar. In the analyses described in this paper, for instance, it was of 0.15 Hz. At this point we go back to the time domain (the sampling frequency is still the original one, 4096 Hz) and make barycentric  and spin-down corrections. 
Due to the Doppler effect the received frequency $f(t)$ is related to the emitted frequency $f_0(t)$ by the well-known relation (valid in the non-relativistic approximation)
 \begin{equation}
f(t)=\frac{1}{2\pi}\frac{d\Phi(t)}{dt}= f_0(t) \left(1+\frac{\vec{v}
\cdot \hat{n}}{c}\right),
\label{eq:fdopp}
\end{equation}
where $\Phi(t)$ is the signal phase, $\vec{v}=\vec{v}_{rev}+\vec{v}_{rot}$ is the detector velocity with respect to the Solar system barycenter (SSB), given by the sum of the Earth's orbital velocity, $\vec{v}_{rev}$, and rotational velocity, $\vec{v}_{rot}$, while $\hat{n}$ is the versor identifying the source position.   
In practice the Doppler effect is efficiently corrected in the time domain by changing the time stamp $t$ of data samples according to 
\begin{equation}
\tau_1=t+\frac{\vec{r(t)}\cdot \hat{n}}{c}=t+\Delta_R
\label{eq:tau1}
\end{equation}
where $\vec{r}$ is the detector position in the SSB.
The correction term $\Delta_R$, called {\it Roemer delay}, amounts up to about $500$ seconds and corresponds to the time taken by a signal traveling at the speed of light to cover the distance between the detector and the SSB. Unlike Eq.~(\ref{eq:fdopp}), Eq.~(\ref{eq:tau1}) does not depend on the frequency, which means that one single correction holds for every frequency.  
In fact there are other smaller relativistic effects that must be taken into account when making barycentric corrections. One is the {\it Einstein delay},  $\Delta_E$, due to the Earth motion and the gravitational redshift at the Earth geocenter due to the solar system bodies, and that amounts to about 2 milliseconds at most. Another effect is the {\it Shapiro delay} $\Delta_S$ which takes into account the deflection of a signal passing near a massive body, and which can be shown to be negligible for CW searches, unless the source line of sight passes very near the Sun's limb. See e.g. \cite{ref:nb_method} for explicit expressions of $\Delta_E$ and $\Delta_S$.
Overall, we can make the full barycentric corrections by introducing the new time variable
\begin{equation}
\tau_1=t+\Delta_R+\Delta_E-\Delta_S
\label{eq:baricorr}
\end{equation}
This transformation corresponds to referring the data collected at the detector site to the SSB, which can be considered an inertial reference frame to a very good approximation. In practice, for the Crab and Vela analyses described in this paper the Shapiro delay can be neglected.
   
For given values of the signal frequency $f$ and frequency derivatives $\dot{f}$, $\ddot{f}$ we could take into account the spin-down in a similar way, by re-scaling time according to
\begin{equation}
\tau_2=t+\frac{\dot{f}}{2f}(t-t_0)^2+\frac{\ddot{f}}{6f}(t-t_0)^3
\label{eq:tau2}
\end{equation}
where $t_0$ is the initial time of the data set and higher order terms can be included if needed.   

Note that re-scaling the time in this way to make the signal monochromatic assumes that the intrinsic phase evolution of the pulsar's GW  signal is described by a Taylor expansion over the entire observation period. Electromagnetic observations of pulsars show that young, rapidly spinning pulsars demonstrate deviations from a Taylor expansion when spinning down, a phenomenon known as \emph{timing noise}.  Given our lack of knowledge of the exact mechanism that might cause the gravitational wave frequency  to deviate from (twice) the observed electromagnetic frequency, we cannot be sure if we should expect the timing noise to be present in the signal we are searching for (see \cite{ref:jones2} for discussion).  However, a study based on the monthly Crab ephemeris data \cite{ref:jodrellweb} has shown that in the Crab, one of the noisiest of pulsars, the effect of timing noise over the duration of the observation period is negligible, producing a mismatch between a Taylor expansion and a signal based on the actual `noisy' time series of less than $1\%$. This is small,
confirming that timing noise is likely to have a negligible effect on our analysis.

\begin{figure*}[!htbp]
\includegraphics[width=10cm]{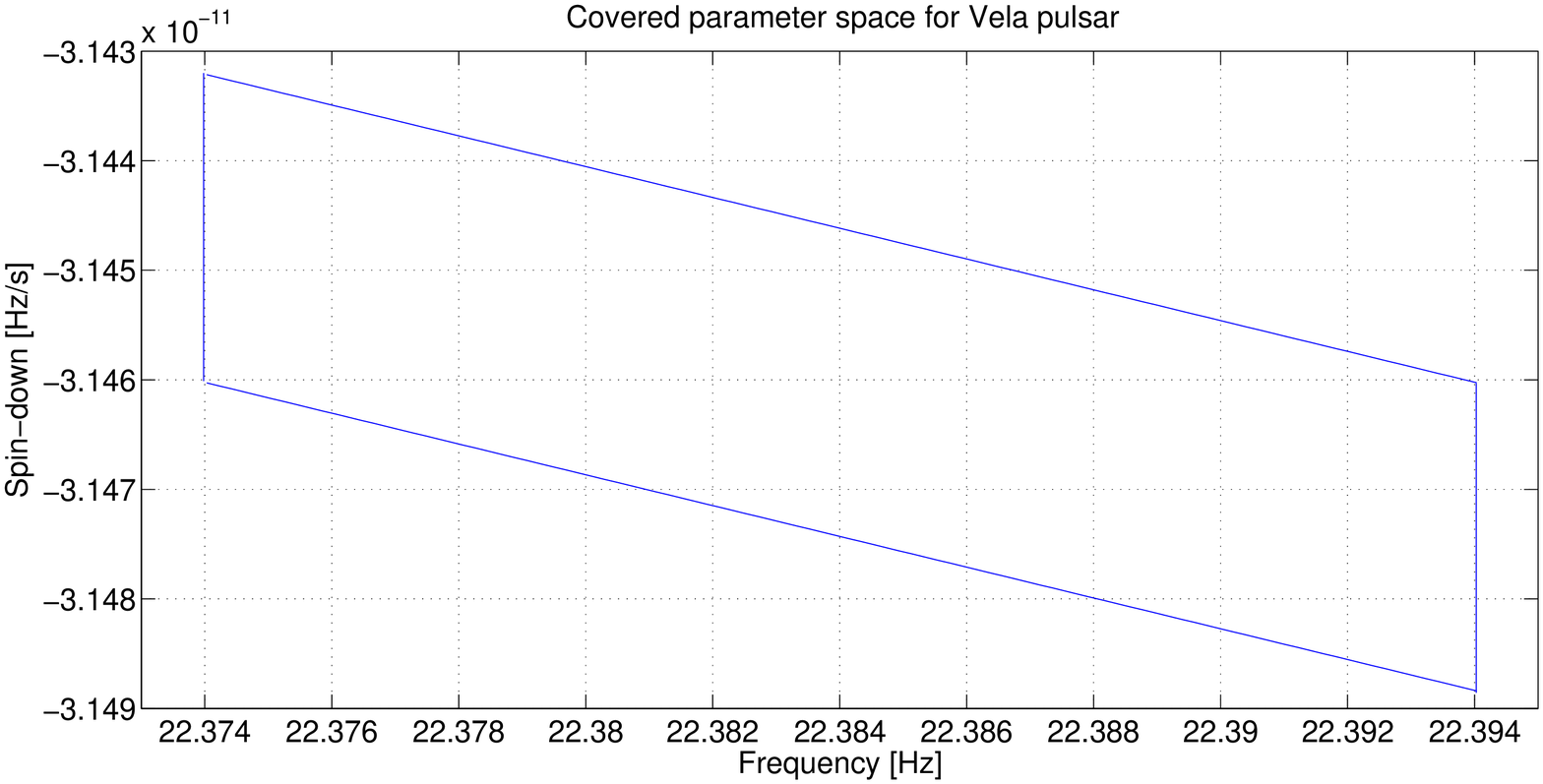}
\includegraphics[width=10cm]{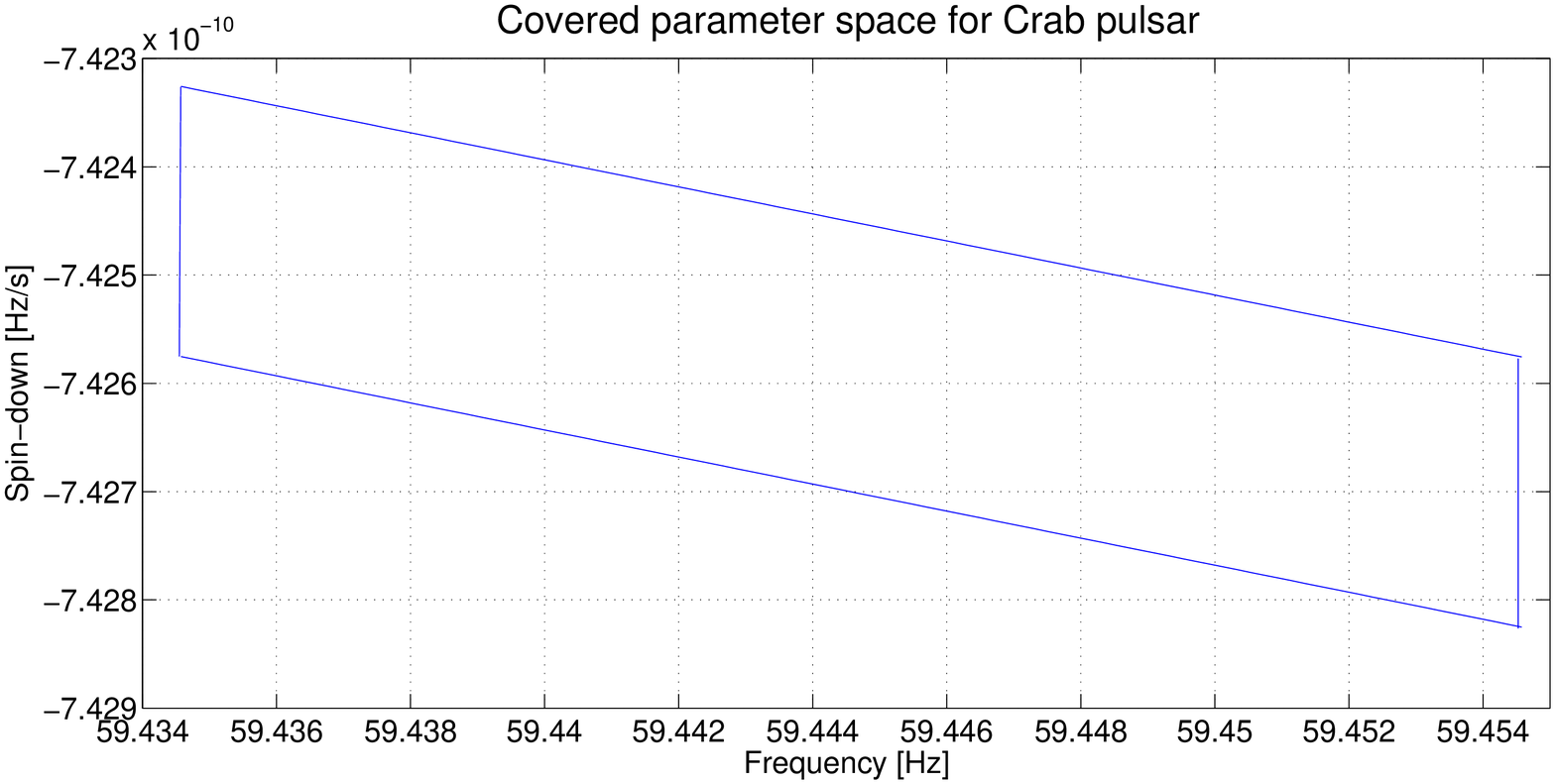}
\caption{The parallelograms delimit the portion of the $f-\dot{f}$ plane covered in the narrow-band search for the Vela (upper plot) and Crab (bottom plot) pulsars. Only one value of $\ddot{f}$ has been considered in the analysis. The total number of points is given in the last column of Tab.~\ref{tab:param}.
\label{fig:ffdot}}
\end{figure*}

In practice the spin-down correction is applied after barycentric corrections, then the time $t$ that appears in Eq.~(\ref{eq:tau2}) is in fact the re-scaled time $\tau_1$ of Eq.~(\ref{eq:baricorr}). 
In this case the correction depends explicitly on the search frequency $f$. The number of frequency bins which cover the range $\Delta f=2f_0\delta$, corresponding to Eq.~(\ref{eq:widef}), is 
\begin{equation}
n_{f}=\left[ \frac{\Delta f}{\delta f}\right ]=\left[ \Delta f \cdot T\right]\approx 6.3\times 10^5\left(\frac{\Delta f}{0.02~{\rm Hz}} \right)\left(\frac{T}{1~{\rm yr}} \right)
\label{eq:nfreq}
\end{equation}
with $\delta f=\frac{1}{T}$ being the frequency spacing, with $T$ the total observation time, and where $\left[ ~\cdot{}~ \right]$ stands for the nearest integer. 
Similarly, we take the width of the first order spin-down range as $\Delta \dot{f}=2|\dot{f}_0|\delta$, and $\Delta \ddot{f}=2|\ddot{f}_0|\delta$ for the second order.    
Let us consider for the moment only the first order term. For a fixed value of $\dot{f}$ we do not want to make an explicit spin-down correction for each value of $f$. So, we fix the value of the frequency in the denominator of Eq.~(\ref{eq:tau2}), for instance at the reference frequency $f_0$, and exploit the fact that the same correction that holds for the pair $(f_0,\dot{f}_0)$ is also valid for all the pairs $(f,\dot{f})$ such that
\begin{equation}
\frac{\dot{f}_0}{f_0}=\frac{\dot{f}}{f}
\end{equation}
In practice, this means that at each frequency $f$ we explore a range of first order spin-down values $[\dot{f}-\frac{\Delta \dot{f}}{2},~\dot{f}+\frac{\Delta \dot{f}}{2}]$, with
\begin{equation}
\dot{f}= \dot{f}_0\frac{f}{f_0}
\end{equation} 
The number of bins in the spin-down term of the first order can be computed by considering the ``natural'' step $\delta \dot{f}=\frac{1}{2T^2}$ and is given by
\begin{align}
n^{(1)}\equiv n_{\dot{f}}=\left[2\Delta \dot{f}\cdot T^2\right]\\ \nonumber
\approx 400\left(\frac{\dot{f}_0}{10^{-10}~{\rm Hz/s}} \right)\left(\frac{\delta}{10^{-3}} \right)\left(\frac{T}{1~{\rm yr}} \right)^2
\label{nsd}
\end{align} 
Following the same reasoning for the second order spin-down we arrive at the following expression for the number of steps that should be taken into account:
\begin{align}
n^{(2)}\equiv n_{\ddot{f}}=\left[\Delta \ddot{f}\cdot T^3\right]\\ \nonumber
\approx 0.6\left(\frac{\ddot{f}_0}{10^{-20}~{\rm Hz/s^2}} \right)\left(\frac{\delta}{10^{-3}} \right)\left(\frac{T}{1~{\rm yr}} \right)^3
\label{eq:nsd2}
\end{align} 
For observation times of the order of a year, reasonable values of second order spin-down, and values of $\delta$ typical of narrow-band searches, $n_{\ddot{f}} < 1$, which means that we do not need to consider more values other than the ``central'' one. 
In Tab.~\ref{tab:ephem} the Crab and Vela EM-inferred positional and rotational parameters are shown. Also estimations of the polarization parameters are given, which are used in the computation of upper limits, see Sec.~\ref{sec:result}.   
\begin{table*}
\caption{\label{tab:ephem}The Crab and Vela reference parameters. $\alpha$ and $\delta$ are the equatorial coordinates. $f_0,\dot{f}_0,\ddot{f}_0$ are the gravitational-wave frequency 
parameters. The reference epoch for the Crab position is MJD 54632, while for the rotational parameters is MJD 55696. The reference epoch for Vela is MJD 53576 both for position and GW 
frequency parameters. The Crab ephemeris has been obtained from a fit of the Jodrell Bank monthly ephemeris \cite{ref:jodrellweb}, while for Vela they have been derived from Hartebeesthoek 
radio telescope observations \cite{ref:buch}. Also estimations of the polarization parameters and their associated uncertainty, obtained from the analysis of Chandra X-ray observations of the 
pulsar wind nebula torus \cite{ref:ng1,ref:ng2}, are given in the last two columns. The analysis presented in \cite{ref:ng1,ref:ng2} does not allow to determine the sense of the star's spin, so values of $\iota$ and $\psi$ corresponding to $\iota \rightarrow 180^{
\circ}-\iota$, $\psi \rightarrow \psi + 180^{\circ}$ are also possible.
However, the upper limits on strain amplitude reported in this paper are not sensitive to the sense of rotation.} 
\begin{tabular}{|c|c|c|c|c|c|c|c|c|c|} \hline
Source & $ \alpha$ [hh:mm:ss] & $ \delta [{\rm deg}] $ & $ f_0 [{\rm Hz}] $ & $ \dot{f}_0 [{\rm Hz/s}] $ & $ \ddot{f}_0 [{\rm Hz/s^2}] $ & $ \iota [{\rm deg}] $ & $ \psi [{\rm deg}] $  \\ \hline \hline
Crab & $ 05:34:31.97 $ & $ 22.0145 $ & $ 59.4448 $ & $ -7.4183\times 10^{-10} $ & $ 2.6307\times 10^{-20} $ & $ 62^o.2\pm 1^o.9 $ & $ 35^o.2\pm 1^o.5$\\ \hline
Vela & $ 08:35:20.61 $ & $  -45.1764 $ & $ 22.3840 $ & $ -3.1460\times 10^{-11} $ & $ 1.2848 \times 10^{-21} $ & $ 63^o.6 \pm 0^o.6 $ & $ 40^o.6\pm 0^o.1$ \\ \hline
\end{tabular}
\end{table*}
In Fig.~\ref{fig:ffdot} the portion of the $f-\dot{f}$ plane actually covered in the Crab and Vela narrow-band search is shown. 

\begin{figure*}[!htbp]
\includegraphics[width=10cm]{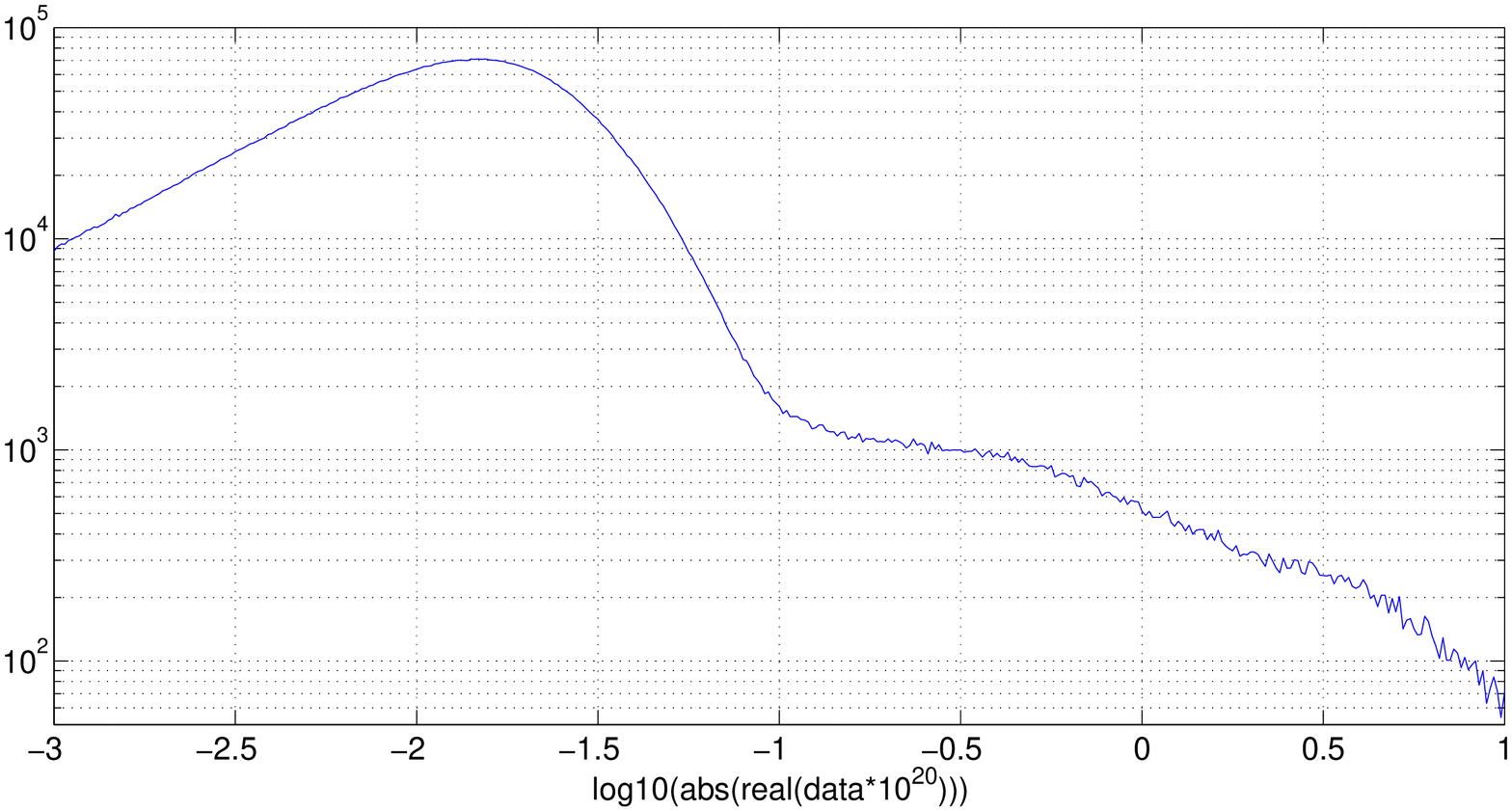}
\includegraphics[width=10cm]{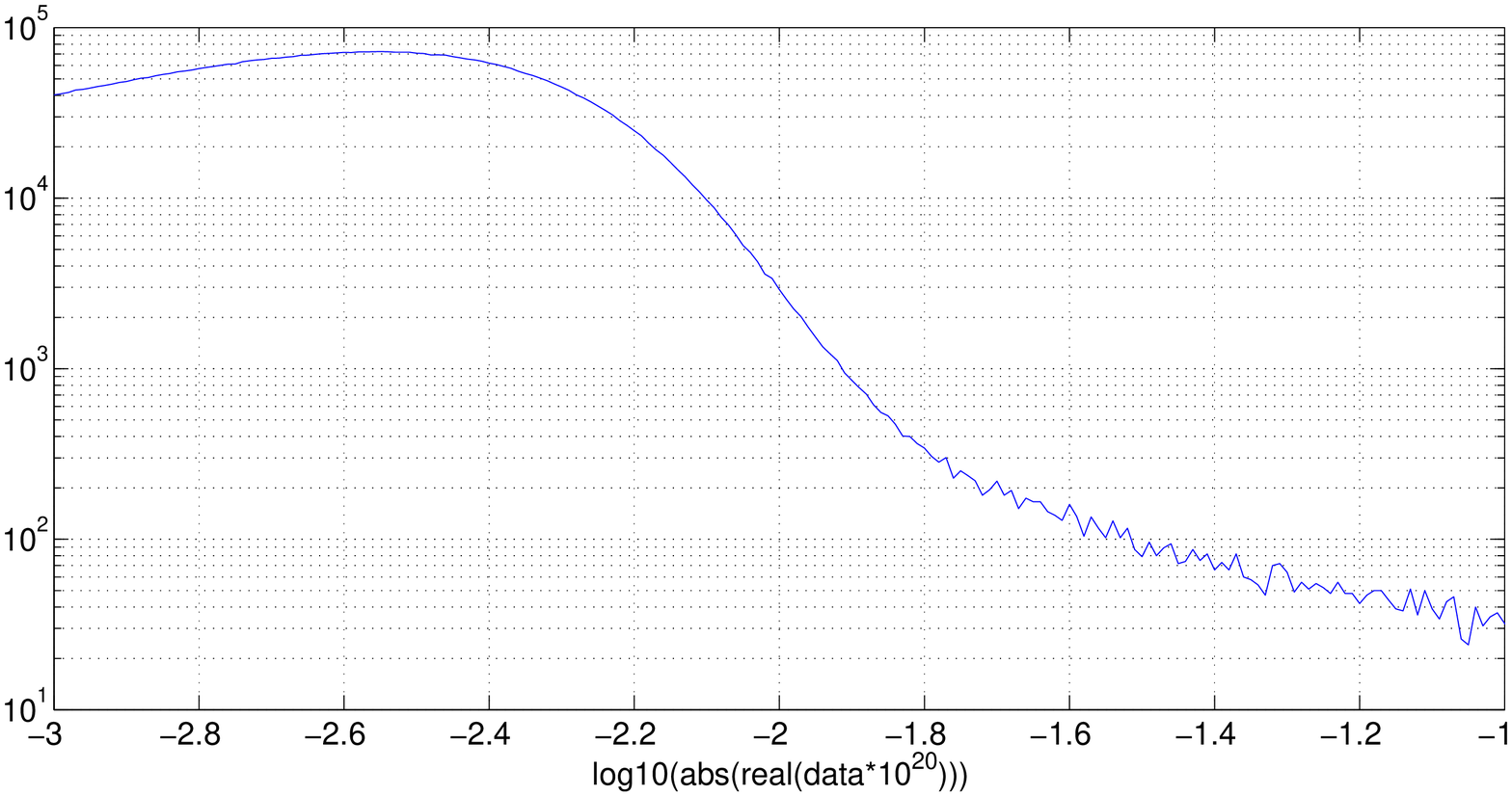}
\caption{Histogram of the logarithm of data amplitude for Vela (top plot) and the Crab (bottom plot), used to select the threshold for the removal of outliers. For Crab the threshold has been put at -1.82, corresponding to an amplitude of $\sim 1.5\times 10^{-21}$, while for Vela a value of -0.92, corresponding to an amplitude of $\sim 1.2\times 10^{-20}$, has been chosen.
\label{fig:datahist}}
\end{figure*}

Once barycentric and spin-down corrections have been done the data are down-sampled to a much lower rate with respect to the original one, 1 Hz in the present case. This strongly reduces the amount of data to be handled in the next step of the analysis. At this stage for a given source we have $n_{\dot{f}}$ corrected time series, one for each value of first order spin-down. For the current analysis we have $n_{\dot{f}}=33$ for Crab and $n_{\dot{f}}=3$ for Vela, see Tab.~\ref{tab:param}) where also other relevant quantities are given. In particular, the total number of points in the frequency/spin-down plane is about $5.28\times 10^6$ for Crab analysis, and $4.8\times 10^5$ for Vela analysis. 
\begin{table*}
\caption{\label{tab:param}Main quantities related to the parameter space of VSR4 narrow-band search. $\Delta f $ is the frequency range; $\delta $ is the width parameter defined in Eq.~(\ref{eq:widef}); $\Delta \dot{f}$ is the first order spin-down range; $n_f$ is the number of frequency bins; $n_{\dot{f}}$ and  $n_{\ddot{f}}$ are, respectively, the number of bins for the first and second order spin-down; $n_{tot}=n_f\cdot  n_{\dot{f}} $ is the total number of points in the parameter space.}
\begin{tabular}{|c|c|c|c|c|c|c|c|}
\hline
Source & $ \Delta f [{\rm Hz}] $ & $ \delta $ & $ \Delta \dot{f} [{\rm Hz/s}] $ & $ n_f $ & $ n_{\dot{f}} $ & $ n_{\ddot{f}} $ & $ n_{tot} $ \\ \hline \hline
Crab & $ 0.02 $ & $ 1.68\times 10^{-4} $ & $ 2.49\times 10^{-13} $ & $ 1.6\times 10^5 $ & $ 33 $ & $ 1 $ & $5.28\times 10^6 $  \\ \hline
Vela & $ 0.02 $ & $  4.47\times 10^{-4} $ & $ 2.81\times 10^{-14} $ & $ 1.6\times 10^5 $ & $ 3 $ & $ 1 $ & $4.80\times 10^5 $  \\ \hline
\end{tabular}
\end{table*}
For each time series we apply a final cleaning step by removing the largest outliers. These are identified by histogramming the logarithm of the absolute value of the data amplitude and choosing a threshold approximately marking the beginning of the non-Gaussian tail of data distribution. In Fig.~\ref{fig:datahist} the histogram of Crab and Vela data amplitude, corresponding to the ``central'' time series, are shown. For Crab a threshold of $1.5\times 10^{-21}$, corresponding to -1.82 in the figure x-axis, has been used to remove outliers, while for Vela a value of $1.2\times 10^{-20}$, corresponding to -0.92, has been used. Correspondingly, the fraction of removed data is of about 1.8$\%$ for Vela and $1.9\%$ for Crab. By applying the Kolmogorov-Smirnov test we have verified that for each pulsar the data distribution of the various time series are fully in agreement thus justifying the use of the same threshold for all of them.  

At this point the detection statistic is computed for every frequency and spin-down value in the explored range. The detection statistic we use is based on the so-called 5-vectors, the same used for pulsar targeted searches \cite{ref:vela_vsr2,ref:vsr4_kp}, and is here briefly described. Once  barycentric and spin-down corrections have been applied, a CW signal with frequency $f_0$ present in the data would be monochromatic, apart from an amplitude and phase sidereal modulation due to the time-varying detector beam pattern functions, and given by Eq.~(\ref{eq:hoft}) with $\omega_0(t)$ constant and equal to $2\pi f_0$. From Eqs.~(\ref{eq:hoft},\ref{eq:Ap},\ref{eq:Ac}) it follows that the signal is completely described by its Fourier components at the 5 angular frequencies $\omega_0,~\omega_0\pm \Omega_{\oplus},~\omega_0\pm 2\Omega_{\oplus}$. This set of 5 complex numbers constitutes the signal {\it 5-vector}. Given a generic time series $g(t)$, the corresponding 5-vector is defined as
\begin{equation}
\mathbf{G}=\int_T g(t)e^{-\imath (\omega_0-\mathbf{k}\Omega_{\oplus})t}dt
\label{eq:def5vect}
\end{equation}
where $\mathbf{k}=[-2,-1,...,2]$ and $T$ is the observation time. Let us indicate with $\mathbf{X}$ the data 5-vector and with $\mathbf{A}^+,~\mathbf{A}^{\times}$ the signal plus and cross 5-vectors, obtained by applying the definition of Eq.~(\ref{eq:def5vect}) to Eqs.~(\ref{eq:Ap},\ref{eq:Ac}). These two last quantities depend only on known parameters and form the signal templates.   
Once the 5-vectors of data and of signal templates have been computed, the two complex numbers
\begin{equation}
\hat{H}_{+/\times}=\frac{\mathbf{X}\cdot \mathbf{A}^{+/\times}}{|\mathbf{A}^{+/\times}|^2}
\label{eq:hphc}
\end{equation}
are built, see \cite{ref:vela_vsr2,ref:fivevect} for more details. They correspond to computing matched filters between the data and the signal templates, and it can be shown, assuming the noise is Gaussian with mean value zero, that they are estimators of the signal plus and cross complex amplitudes $H_0e^{\imath \Phi_0}H_+,~H_0e^{\imath \Phi_0}H_{\times}$. These estimators are used to build the detection statistic 
\begin{equation}
\mathcal{S}=|\mathbf{A}^{+}|^4|\hat{H}_{+}|^2+|\mathbf{A}^{\times}|^4|\hat{H}_{\times}|^2
\label{eq:detstat}
\end{equation}
The maximum of the detection statistic over the searched parameter space, $\mathcal{S}_{max}$, is determined. This is the loudest candidate and is identified by a triple $(\mathcal{S}_{max}, f_{\mathcal{S}_{max}}, \dot{f}_{\mathcal{S}_{max}})$. This means we need to compute $n_{tot}=n_f\cdot  n_{\dot{f}} $ values of the detection statistic. 

The maximum value $\mathcal{S}_{max}$ is used to assess detection significance by computing the $p$-value, that is the probability that a value of the detection statistic equal, or larger, than $\mathcal{S}_{max}$ can be obtained in the absence of any signal. It implies the need to compute the noise-only distribution of the detection statistic. This is a multi-dimensional probability distribution (with dimension $n_{tot}$) which would be difficult to compute and to handle. In practice, as discussed in \cite{ref:nb_method}, the $p$-value is computed considering the single-trial noise probability distribution, that is the same that would be used for a targeted search, and choosing a suitable threshold on it to discriminate between interesting (that is, deserving a deeper study) and not interesting candidates. As shown in \cite{ref:nb_method}, by setting an overall $p$-value $p_0$ (over the full multi-dimensional distribution) the corresponding threshold on the single-trial distribution is $p_{thr}=1-\left(1-p_{0}\right)^{\frac{1}{n_{tot}}}$. 
In our case, by setting, e.g., $p_0=0.01$ the resulting $p_{thr}$ would be of the order of $10^{-8}$. In principle, we would like to generate the noise-only probability distribution by computing the detection statistic at several ``off-source'' frequency bins, that is frequencies near but outside the explored range, where we are assuming the signal could be. This is what is typically done in targeted searches \cite{ref:vsr4_kp}. In the present analysis, however, in order to appreciate $p$-values of the order of $p_{thr}$ we should consider of the order of $10^8$ off-source frequency bins. This is computationally impractical. For this reason we use the theoretical distribution, which assumes the noise is Gaussian, given in \cite{ref:nb_method}:
\begin{equation}
f(\mathcal{S})=\frac{e^{-\frac{\mathcal{S}}{\sigma^2_X|\mathbf{A}^{\times}|^2}}-e^{-\frac{\mathcal{S}}{\sigma^2_X|\mathbf{A}^{+}|^2}}}{\sigma^2_X\left(|\mathbf{A}^{\times}|^2-
|\mathbf{A}^{+}|^2\right)}
\label{eq:pdfS}
\end{equation}
where $\sigma^2_X=\sigma^2\cdot T$, with $\sigma^2$ being the noise variance.

If $S_{max}$ is compatible with noise at a given confidence level, e.g. 1$\%$, we compute an upper limit on signal strength, using the same method described in \cite{ref:vsr4_kp}, otherwise in case of detection signal parameters are estimated through suitable combinations of the real and imaginary parts of $\hat{H}_{+}$ and $\hat{H}_{\times}$, as explained in \cite{ref:fivevect}.   

\section{\label{sec:result}Results}
\begin{figure*}[!htbp]
\includegraphics[width=1.0\textwidth]{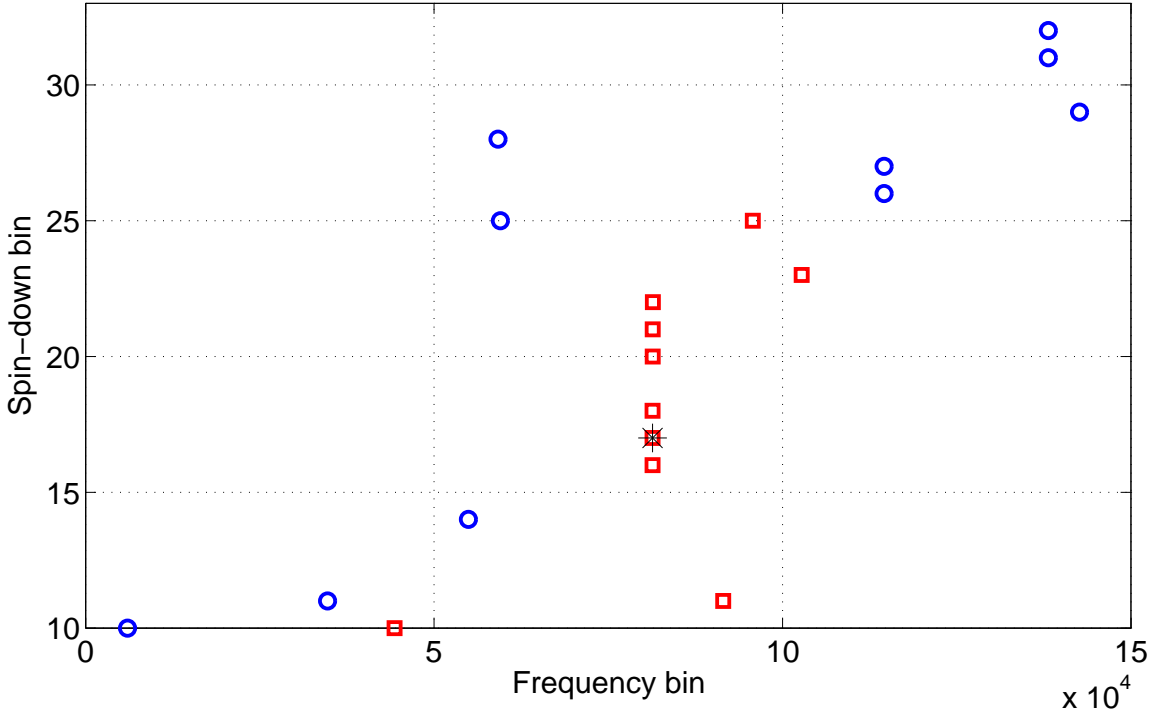}
\caption{Distribution in the frequency/spin-down plane of the top ten candidates obtained in the actual analysis (blue circles) and obtained after the addition to the data of a simulated signal with amplitude $h_0=4.06\times 10^{-25}$, such that the resulting loudest candidate is approximately as loud as the one obtained in the actual analysis (red squares). The black star identify the injection. Frequency and spin-down are expressed as number of bins from the beginning of the corresponding intervals. Six red squares appear to be at the same frequency, but this is just a visual effect due to the large range of frequency covered by the x-axis. In fact, they are very near but spread around the injection values.
\label{fig:candclust}}
\end{figure*}

Analysis results are summarized in Tab.~\ref{tab:results}.
The data, both for the Vela and Crab searches, are compatible with noise. In particular, we find $p$-values equal to 0.33 and 0.013 for Vela and the Crab respectively, larger than $p_0=0.01$ chosen to identify interesting candidates. For Crab, however, the obtained $p$-value is very near to the chosen threshold. Altough a reasonable choice of $p_0$ is rather arbitrary (we could have for instance chosen $10^{-3}$), we have decided to study in some detail the candidate's properties. In particular, we have considered the distribution of the top ten candidates in the frequency/spin-down plane. They appear to be randomly distributed, without the clustering that we would expect in presence of a signal. To verify this hypothesis we have added a simulated signal to the data, with the same parameters as the Crab, but with a slightly different frequency and spin-down, and with an amplitude $h_0\simeq 4.1\times 10^{-25}$ such that the resulting loudest candidate has a value of the detection statistic approximately equal to the loudest candidate of the actual analysis. We have then run the full analysis on this new data set and looked again at the top ten candidate distribution in the frequency/spin-down plane. In this case we indeed observe a clustering, with 4 out of 10 candidates having a frequency within $\pm 2$ bins of the injected value, and 4 out of 10 candidates having a spin-down within $\pm 3$ bins of the injected value. Overall, 5-6 of the top ten candidates appear to be due to the injected signal. Top ten candidates distribution in the frequency/spin-down plane is shown in Fig.~\ref{fig:candclust} for both cases. The results of this test make us more confident in declaring a non-detection also in the Crab analysis. 

We have computed 95$\%$ confidence level upper limits, both assuming uniform priors on the polarization parameters, $\psi$ and $\cos{\iota}$, and ``restricted" priors described by a Gaussian distribution with mean value and standard deviation given in Tab.~\ref{tab:ephem}. In fact the upper limits we compute are obtained from the posterior distribution of the signal strain amplitude, conditioned to the observed value of the detection statistic, as described in \cite{ref:vsr4_kp}. 
\begin{table*}
\caption{\label{tab:results}Summary of the analysis results. The $p$-value is reported in the second column; $ h_{UL}^{unif} $ is the experimental upper limit assuming uniform priors for the polarization parameters; $ h_{UL}^{restr} $ is the experimental upper limit assuming ``restricted'' priors for the polarization parameters, given by a Gaussian distribution with mean value and standard deviation given in Tab.~\ref{tab:ephem}; $ h_{sd} $ is the spin-down limit, computed through Eq.~(\ref{eq:h0sd}); $\epsilon_{UL}$ is the upper limit on the ellipticity; $Q_{22,UL}$ is the upper limit on the mass quadrupole moment; $h_{UL}/h_{sd}$ is the ratio between the upper limit and the spin-down limit on signal strain amplitude; $ \dot{E}_{UL}/\dot{E}_{sd} $ is the upper limit on the fraction of spin-down energy due to the emission of gravitational waves (the values among parentheses refer to the restricted case).}
\begin{tabular}{|c|c|c|c|c|c|c|c|c|}
\hline
Source & $p$-value & $ h_{UL}^{unif} $ & $ h_{UL}^{restr} $ & $ h_{sd} $ & $ \epsilon_{UL} \times 10^4 $ & $ Q_{22,UL}\times 10^{-34}~[{\rm kg~m^2}] $ & $ h_{UL}/h_{sd} $ & $ \dot{E}_{UL}/\dot{E}_{sd} $  \\ \hline \hline
Vela & $ 0.33 $ & $ 3.2\times 10^{-24} $ & $ 3.3\times 10^{-24} $ & $ 3.3\times 10^{-24} $ & $ 17.6 (19.1) $ & $ 13.4 (13.9) $ & $ 0.97 (1) $ & $ 0.94 (1) $  \\ \hline
Crab & $ 0.013 $ & $ 7.0\times 10^{-25} $ & $ 6.9\times 10^{-25} $ & $ 1.4\times 10^{-24} $ & $ 3.8 (3.7) $ 
& $ 2.9 (2.8) $ & $ 0.50 ( 0.49) $ & $ 0.25 (0.24) $  \\ \hline
\end{tabular}
\end{table*}
For Vela the upper limits are very similar to the spin-down limit. This does not allow us to significantly constrain the fraction of spin-down energy due to the emission of gravitational waves. In the case of the Crab pulsar the upper limits on signal strain amplitude are about 2 times below the spin-down limit, with a corresponding constraint of about 25$\%$ on the fraction of spin-down energy due to gravitational waves. The upper limit on signal strain amplitude can be converted, via Eq.~(\ref{eq:eps_sd}), into an upper limit on star ellipticity of about $3.7\times 10^{-4}$, assuming the neutron star moment of inertia is equal to the canonical value of $10^{38}\rm{kg~m^2}$. The upper limits on ellipticity are comparable to the maximum value foreseen by some ``exotic'' equation of state for neutron star matter \cite{ref:owen,ref:hask}, but are much larger than the maximum value predicted for a ``standard'' equations of state. In Tab.~\ref{tab:results} upper limits on the quadrupole mass moment, which are independent on the uncertain value of the star moment of inertia, are also given. An uncertainty of about 8$\%$ is associated with the upper limit on strain amplitude. This has been estimated as the square root of the quadratic sum of the calibration error which, as discussed in Sec.~\ref{sec:data}, amounts to 7.5$\%$, and of the error associated to the finite size of the simulation used to compute the upper limits, which is about 3$\%$. 

While this is the first time a narrow-band search has been carried out for the Vela pulsar, our results for the Crab represent an improvement with respect to the LIGO/S5 search described in \cite{ref:s5_nb}. Overall, this is the first time the spin-down limit has been significantly overcome in any narrow-band search. On the other hand, our upper limits are clearly worse than those found in Vela and Crab targeted searches \cite{ref:vsr4_kp}. As explained in \cite{ref:nb_method} this is what we expect as a consequence of the lower sensitivity of narrow-band searches, due to the volume of parameter space that is explored. Also, the use of VSR4 data alone for the Crab search contributes, even if to a lesser extent, to reduce the search sensitivity with respect to \cite{ref:vsr4_kp} where data from Virgo VSR2, VSR4 and LIGO S6 were used.     

\section{\label{sec:valid} Validation tests}
We have performed validation tests of the analysis pipeline both injecting software simulated signals in Gaussian noise and performing a narrow-band search around some of the hardware injected signals in VSR4 data. The first kind of test has been discussed in \cite{ref:nb_method}. Here we focus attention on the second kind of test. 

\begin{figure*}[!htbp]
\includegraphics[width=10cm]{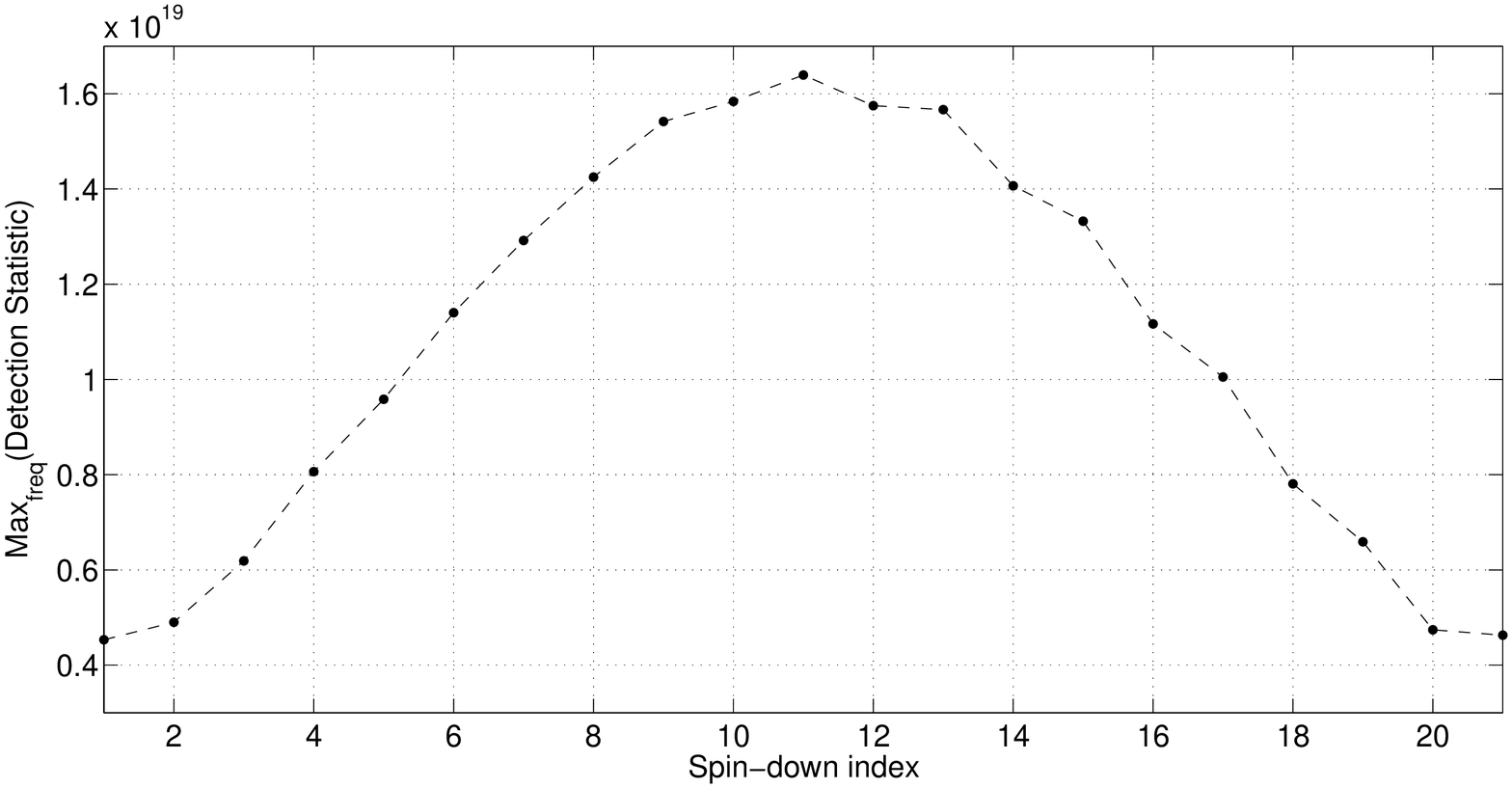}
\includegraphics[width=10cm]{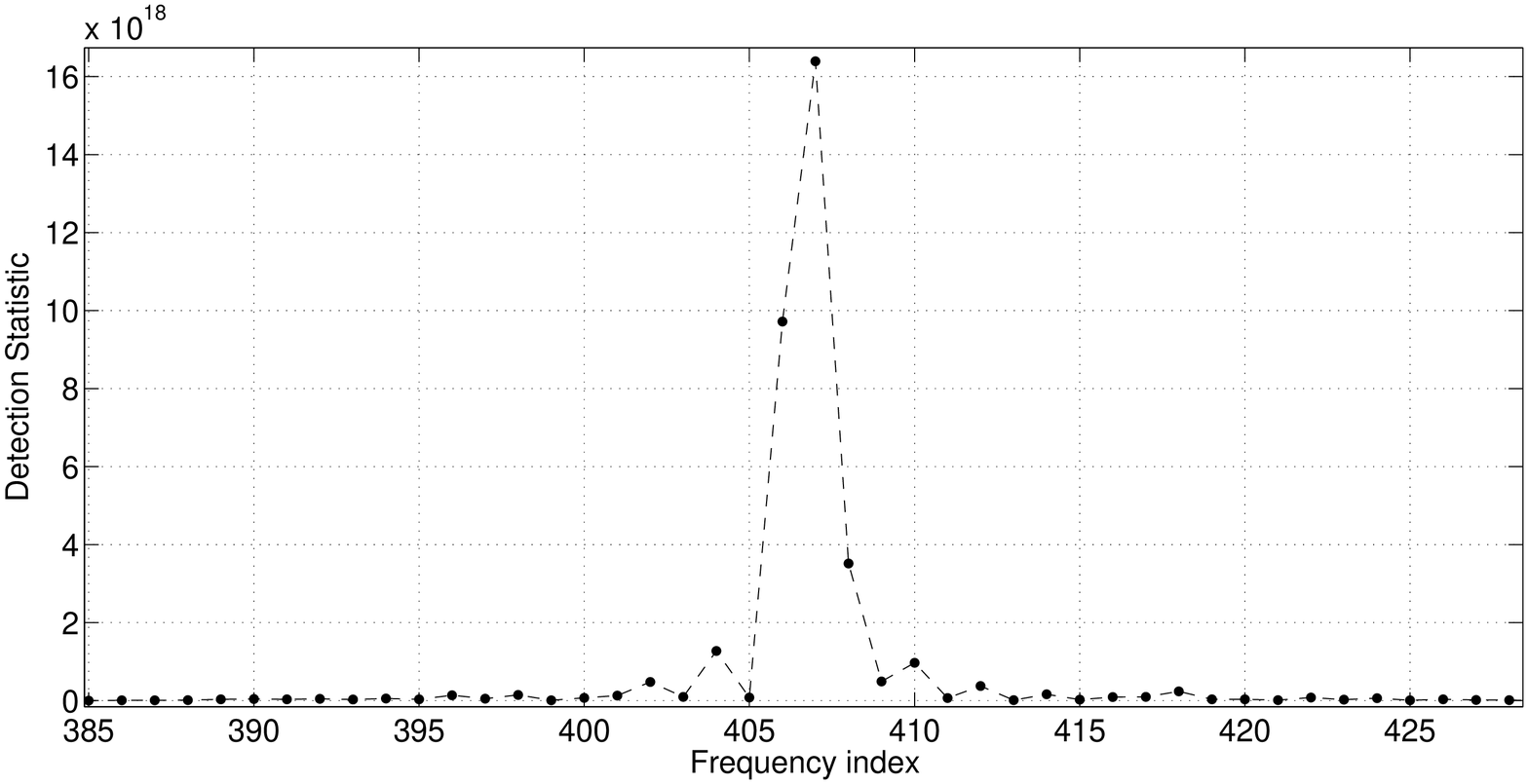}
\caption{Tests with hardware injection {\it Pulsar3}. Upper plot: maximum of the detection statistic (maximized over frequency) as a function of the spin-down index (which goes from 1 to 21). Bottom plot: value of the detection statistic as a function of the frequency bin (which goes from 1 to 814), computed taking the spin-down of the loudest candidate. The plots show that the frequency and spin-down of the loudest candidate are the nearest to the true values.
\label{fig:hi_p3}}
\end{figure*}

For the entire duration of VSR4 run, 10 simulated CW signals have been injected in the Virgo detector by sending the appropriate excitations to the coils used to control mirror position. These signals were characterized by various amplitudes, frequency, spin-down and polarization parameters, and corresponded to sources with various locations in the sky. In particular here we have considered injections named {\it Pulsar3, Pulsar5, Pulsar8} with parameters of signal amplitude $h_0$, frequency $f_0$, spin-down $\dot{f}_0$, position $(\alpha_0,\delta_0)$, ratio between the polarization ellipse semi-minor and semi-major axes $\eta_0$ and wave polarization angle $\psi_0$, given in Tab.~\ref{tab:hi_param}. 
\begin{table*}
\caption{\label{tab:hi_param}Main parameters of the hardware injections used to test the narrow-band search pipeline. $h_0$ is the signal amplitude; $f_0$ is the signal frequency; $\dot{f}_0$ is the spin-down, $(\alpha_0,~\delta_0)$ is the source position in equatorial coordinates, $\eta_0$ is the ratio between the polarization ellipse semi-minor and semi-major axes, $\psi_0$ is the wave polarization angle.}
\begin{tabular}{|c|c|c|c|c|c|c|c|}
\hline
Name & $ h_0 $ & $ f _0[{\rm Hz}] $ & $ \dot{f}_0 [{\rm Hz/s}] $ & $ \alpha_0 [{\rm deg}] $ & $ \delta_0 [{\rm deg}] $ & $ \eta_0 $ & $ \psi_0~ [{\rm deg}] $ \\ \hline \hline
Pulsar3 & $ 8.296\times 10^{-24} $ & $ 108.857159396 $ & $ -1.46\times 10^{-17} $ & $ 178.372574 $ & $ -33.436602 $ & $ 0.1600 $ & $ 25.439 $  \\ \hline
Pulsar5 & $ 3.703\times 10^{-24} $ & $ 52.808324359 $ & $  -4.03\times 10^{-18} $ & $ 302.626641 $ & $ -83.8391399 $  & $ -0.7625 $ & $ -20.853 $ \\ \hline
Pulsar8 & $ 8.067\times 10^{-24} $ & $ 192.237058812 $ & $ -8.65\times 10^{-9} $ & $ 351.389582 $ & $ -33.4185168 $ & $ -0.1470 $ & $ 9.7673 $ \\ \hline
\end{tabular}
\end{table*}
For each of the hardware injections we have performed a narrow-band search over a frequency range of $10^{-4}$ Hz and over a spin-down range of $1.585\times 10^{-13}$ Hz/s around the signal injected values. The grid in frequency and spin-down has been built in such a way that each center bin corresponds to a frequency given by an integer number of bins. This means that the true values do not coincide with a bin center. The explored frequency range is covered by 814 bins while the spin-down range is covered by 21 bins. For each hardware injection the frequency and spin-down corresponding to the loudest candidate have been selected and compared to the true signal values. Moreover, the amplitude and polarization parameters of the injected signals have been estimated and compared to the actual values and to the values that would have been found in a targeted search. In Tab.~\ref{tab:hi_res1} and \ref{tab:hi_res2} test results are summarized.
\begin{table*}
\caption{\label{tab:hi_res1}Tests with hardware injections: frequency and spin-down recovery. For each hardware injection the estimated frequency and spin-down, $f$ and $\dot{f}$, and the associated errors $\epsilon_f$ and $\epsilon_{\dot{f}}$, expressed in number of bins, are given.}
\begin{tabular}{|c|c|c|c|c|}
\hline
Name & $ f [{\rm Hz}] $ & $ \epsilon_f $ & $ \dot{f} [{\rm Hz/s}]  $ & $ \epsilon_{\dot{f}} $  \\ \hline \hline
Pulsar3 & $ 108.857159411 $ & $ 0.12 $ & $ 0 $ & $ 0.002 $  \\ \hline
Pulsar5 & $ 52.808324371 $ & $ 0.10$ & $  0 $ & $ 0.0005 $  \\ \hline
Pulsar8 & $ 192.237058796 $ & $ -0.13$ & $ -8.6500021\times 10^{-9} $ & $ -0.272 $  \\ \hline
\end{tabular}
\end{table*}
In particular, in Tab.~\ref{tab:hi_res1} we report the estimated frequency $f$, the error with respect to the true value $\epsilon_f$, measured in number of frequency bins, the estimated spin-down value $ \dot{f}$ and its error $ \epsilon_{\dot{f}}$ in number of spin-down bins. 
From Tab.~\ref{tab:hi_res1} we see that, within the discretization error, both the frequency and the spin-down are correctly recovered for all the hardware injections. This can be also seen in Fig.~\ref{fig:hi_p3} where two plots, referring to {\it Pulsar3}, are shown. 
The upper plot represents the maximum of the detection statistic (maximized over frequency) as a function of the spin-down index, which goes from 1 to 21. The maximum of the curve is at index number 11 which is the nearest to the injected value. Note that given the very small spin-down value of {\it Pulsar3} and {\it Pulsar5}, much smaller than the spin-down bin width, the searched range covers both negative and positive values of $\dot{f}$. The lower plot shows the value of the detection statistic as a function of the frequency bin, which goes from 1 to 814, computed taking the spin-down of the loudest candidate. As before the maximum is found in correspondence of the bin nearest to the injected value of the frequency.  
\begin{table*}
\caption{\label{tab:hi_res2}Tests with hardware injections: amplitude and polarization parameter recovery. For each hardware injection we report the ratio between the estimated and injected signal amplitude, $h/h_0$, the relative error on $\eta$ normalized to its range of variation (2), $\epsilon_{\eta}$, and the relative error on ${\psi}$ normalized to its range of variation (90 degrees), $\epsilon_{\psi}$. The numbers in parentheses refer to a targeted search for the same signals.}
\begin{tabular}{|c|c|c|c|}
\hline
Name & $ h/h_0 $ & $ \epsilon_{\eta} $ & $ \epsilon_{\psi} $ \\ \hline \hline
Pulsar3 & $ 1.016 (0.992) $ & $ 0.0027 (0.0021) $ & $ -0.0079 (-0.0081) $   \\ \hline
Pulsar5 & $ 1.023 (1.002) $ & $  0.0080 (0.0089) $ & $ 0.1196 (0.0104) $   \\ \hline
Pulsar8 & $ 0.871 (0.975) $ & $ 0.0278 (0.0038) $ & $ 0.0007 (0.0064) $  \\ \hline
\end{tabular}
\end{table*}
In Tab.~\ref{tab:hi_res2} the error in the estimation of amplitude and polarization parameters of hardware injections is given. In particular, in column 2 the ratio between the found and the injected amplitude is reported, in column 3 the fractional error on $\eta$, that is $\epsilon_{\eta}=\frac{\eta-\eta_0}{2}$, and in column 4 the fractional error on $\psi$, $\epsilon_{\psi}=\frac{\psi-\psi_{0}}{90}$. Values in parentheses refer to a targeted search for the same injections. 
Parameters are generally well recovered, with an accuracy just slightly worse with respect to the targeted search case. This is due to the small error in the estimation of the signal frequency in a narrow-band search, consequence of the finite size of the grid step. In fact, looking for instance at the amplitude estimation, we can note that for {\it Pulsar8}, for which the frequency error is the largest, the loss with respect to the targeted search is the biggest, as expected.   

\section{\label{sec:disc}Conclusions}
Targeted searches for continuous gravitational wave signals assume a strict correlation between the gravitational wave frequency and the star rotation frequency. For instance, for a neutron star non-axisymmetric with respect to the rotation axis the gravitational wave frequency is exactly two times the rotation rate. However, it is questionable that such strict a correlation is always valid or that is maintained over long times, and various mechanisms have been proposed that could produce a mismatch. For this reason it is important to have in place an analysis procedure robust with respect to deviations from the standard assumption made in targeted searches and then able to perform a narrow-band search for CW signals over a range of frequency and spin-down values. 
In this paper we have presented results of a narrow-band search for CW from the Crab and Vela pulsars in the data of Virgo VSR4 run. In both cases the data appear to be fully compatible with noise, and we have set 95$\%$ confidence level upper limits on strain amplitude assuming both uniform and restricted priors on polarization parameters. The upper limits are, respectively, comparable (for Vela) and a factor of 2 below (for the Crab) to the corresponding spin-down limits and while this is the first time a narrow-band search has been done for the Vela pulsar, for the Crab our results significantly improve with respect to past analyses. 
As expected, the narrow-band search upper limits are worse than those established in targeted searches of the same sources. This analysis method will be applied to narrow-band searches of CW signals from several potentially interesting pulsars in data of advanced Virgo and LIGO detectors, which will start to collect science data at the end of 2015, and are expect to reach their target sensitivity, about one order of magnitude better than first generation detectors, around 2018 \cite{ref:obscen}.

\begin{acknowledgments}
The authors gratefully acknowledge the support of the United States
National Science Foundation for the construction and operation of the
LIGO Laboratory, the Science and Technology Facilities Council of the
United Kingdom, the Max-Planck-Society, and the State of
Niedersachsen/Germany for support of the construction and operation of
the GEO600 detector, and the Italian Istituto Nazionale di Fisica
Nucleare and the French Centre National de la Recherche Scientifique
for the construction and operation of the Virgo detector. The authors
also gratefully acknowledge the support of the research by these
agencies and by the Australian Research Council, 
the International Science Linkages program of the Commonwealth of Australia,
the Council of Scientific and Industrial Research of India, 
the Istituto Nazionale di Fisica Nucleare of Italy, 
the Spanish Ministerio de Econom\'ia y Competitividad,
the Conselleria d'Economia Hisenda i Innovaci\'o of the
Govern de les Illes Balears, the Foundation for Fundamental Research
on Matter supported by the Netherlands Organisation for Scientific Research, 
the Polish Ministry of Science and Higher Education, the FOCUS
Programme of Foundation for Polish Science,
the Royal Society, the Scottish Funding Council, the
Scottish Universities Physics Alliance, The National Aeronautics and
Space Administration, OTKA of Hungary, the Lyon Institute of Origins (LIO),
the National Research Foundation of Korea,
Industry Canada and the Province of Ontario through the Ministry of Economic Development and Innovation, the National Science and Engineering Research Council Canada,the Carnegie Trust, the Leverhulme Trust, the David and Lucile Packard Foundation, the Research Corporation, and the Alfred P. Sloan Foundation.
\end{acknowledgments}

\clearpage

\bibliography{narrowband_observational}

\end{document}